\documentclass[journal,12pt,onecolumn]{IEEEtran}
\usepackage[letterpaper, left=1in, right=1in, bottom=1in, top=0.75in]{geometry}
\usepackage{caption}

\usepackage{anyfontsize}

\usepackage{cite}
\usepackage{graphicx}
\usepackage{psfrag}
\usepackage{subfigure}
\usepackage{stfloats}
\usepackage{amsmath}
\usepackage{epsfig}
\usepackage{amsfonts}
\usepackage{dsfont}
\usepackage{color}
\usepackage{physics}
\usepackage{tikz,pgfplots}
\usetikzlibrary{shapes,arrows}
\usepackage[ruled,vlined]{algorithm2e}

\DeclareMathOperator*{\argmax}{\arg\max}
\DeclareMathOperator*{\argmin}{\arg\min}
\pgfplotsset{compat=1.14}

\usepackage{setspace}
\begin{document}
\fontsize{12.2pt}{14.4pt}\selectfont

\title{Encoders and Decoders for Quantum Expander Codes Using Machine Learning}


\author{Sathwik Chadaga, Mridul Agarwal, and Vaneet Aggarwal \thanks{S. Chadaga is with IIT Madras, Chennai, India, email: chadagasathwik@gmail.com. He was with Purdue University when this work was done. M. Agarwal and V. Aggarwal are with Purdue University, West Lafayette IN, USA, email: \{agarw180,vaneet\}@purdue.edu. }}


%Telephone: (800) 555--1212, Fax: (888) 555--1212}
% \author{\IEEEauthorblockN{Author 1, Author 2\IEEEauthorrefmark{2}}
% \IEEEauthorblockA{Department of Electrical Engineering\\
% Indian Institute of Technology Madras, Chennai, India\\
% Email: email@ee.iitm.ac.in}
% \IEEEauthorblockA{\IEEEauthorrefmark{2}Extra affiliations}
% }


\maketitle

%Introduction of ISI complicates the receiver structure but this can be eliminated by precoding techniques to preprocess data.
% Finally, we simulate the performance of the system with precoding technique and compare its performance with the Nyquist system in single carrier and OFDM.
\begin{abstract}
Quantum key distribution (QKD) allows two distant parties to share encryption keys with security based on laws of quantum mechanics. In order to share the keys, the quantum bits have to be transmitted from the sender to the receiver over a noisy quantum channel. In order to transmit this information, efficient encoders and decoders need to be designed. However, large-scale design of quantum encoders and decoders have to depend on the channel characteristics and require look-up tables which require memory that is exponential in the number of qubits. In order to alleviate that, this paper aims to design the quantum encoders and decoders for expander codes by adapting techniques from machine learning including reinforcement learning and neural networks to the quantum domain. The proposed quantum decoder trains a neural network which is trained using the maximum aposteriori error for the syndromes, eliminating the use of large lookup tables. The quantum encoder uses deep Q-learning based techniques to optimize the generator matrices in the quantum Calderbank-Shor-Steane (CSS) codes. The evaluation results demonstrate improved performance of the proposed quantum encoder and decoder designs as compared to the quantum expander codes.   
\end{abstract}
%\newpage

% \fi
\IEEEpeerreviewmaketitle

\tikzset{
  block/.style    = {draw, rectangle, minimum height = 3em, minimum width = 3em},
  sum/.style      = {draw, circle, node distance = 1cm},
  input/.style    = {coordinate}
}

\section{Introduction}
\subsection{Overview}
Quantum codes are used to transmit quantum bits (qubits) efficiently over noisy quantum channels. A formulation of quantum encoders have been proposed in  \cite{calderbank1998quantum,calderbank1997quantum,gottesman1997stabilizer}. These encoder designs are based on the minimum distance of the codes. Further, to decode the noisy qubit, typical algorithms include syndrome lookup table. The lookup table is based on the lower error weight corresponding to the syndrome \cite{gottesman1997stabilizer}. The key issue with the encoder design is that it does not take into account the characteristics of the noise in the channel and the key issue with the decoder is that it considers lower weight errors as more likely and the memory requirements to store such a table are high. This paper aims to alleviate these by proposing a machine learning based methodology for design of quantum encoders and decoders.

%Different algorithms are used to correct the errors applied by the channel. One such simple algorithm is to generate a  syndrome lookup table and lookup this table to correct the errors.  Other algorithms have been proposed that search for a low weight error that minimizes the weight of the measured syndrome. One of the disadvantages of these algorithms is that they inherently assume that the low weight errors have higher probability and also assume that different quantum error types have equal probabilities. Also the memory used by the lookup table method of error correction increases exponentially with the dimensions of the quantum code. Hence, we propose an error correction method in which a neural network is trained to predict the channel errors. This method does not require exponential memory space unlike the syndrome lookup table method. This method also learns to perform well under asymmetric channels where the error probabilities of error types need not be equal. We use the performance measure of the neural network as reward to learn and construct new codes using a deep Q-network (DQN) agent. Since the neural network performs well for asymmetrical channels, the agent has the potential to learn new codes that perform better under asymmetrical channels.

\subsection{Related Work}

{\bf Quantum Codes: } The Calderbank-Shor-Steane (CSS) code construction is a special case of stabilizer code construction of quantum error
correcting codes (QECC), and takes a classical binary code that is
self-orthogonal with respect to a certain symplectic inner product to produce a quantum code with minimum distance
determined by the classical code (for more details see \cite{calderbank1998quantum,calderbank1997quantum,gottesman1997stabilizer}). Even though many codes achieving the minimum distance have been found \cite{grassl2016quantum,aggarwal2008boolean,huber2019quantum}, codes achieving minimum distance may not be optimal for general channels. To understand this, let us assume that the channel errors only consists of single bit-flip errors. Then, one qubit can be encoded into 3 qubits rather than the 5 qubits that are required for codes to correct any possible single qubit erros. Recently, large classes of codes based on expander codes \cite{leverrier2015quantum,fawzi2018constant}, LDPC codes \cite{djordjevic2008quantum,xie2018design}, and turbo codes \cite{izhar2018quantum,xiao2017construction} have been studied. One of the key challenge in efficient quantum code designs for large parameters is the storage and computation of the translation from the syndrome to the corrected error. For an $[[n,k]]$ quantum code, this is a mapping from each of $2^{n-k}$ syndromes to $\{I,X,Y,Z\}^n$ error vector, where $I$ is identity matrix and $X, Y, Z$ are Pauli matrices. For large $n-k$, this storage is infeasible. To alleviate this, a small-set-flip algorithm has been proposed in \cite{leverrier2015quantum}. Even though the method is linear in the number of qubits, the complexity is exponential in the weight of the generators. In order to not have such bottleneck, this paper provides a machine learning based approach for encoder and decoder designs that has low storage and computation requirements for the decoder. 

{\bf Machine Learning for Code Generation: } In classical communication systems, machine learning based transceivers have been studied. The authors of \cite{gruber2017deep} studied the decoder design using an artificial neural network and showed that their decoder could approach maximum a posteriori (MAP) performance for short codes. The design of encoders with machine learning approaches has been studied in \cite{huang2019ai}, where a constructor-evaluator framework is proposed to design error correction codes. Our work aims to have joint encoder and decoder designs for quantum codes, where reinforcement learning \cite{sutton2018reinforcement} is used for encoder where the decoding happens with a neural network based decoder. 

{\bf Machine Learning for Quantum Codes: } Reinforcement learning has been used for topological quantum encoder designs \cite{nautrup2018optimizing}. In contrast, this paper considers the quantum expander codes framework considered in \cite{leverrier2015quantum}. For the quantum expander codes, in addition to finding an efficient encoder, the decoder design is also essential since the known approaches are of high complexity. Our work aims to find joint encoder and decoder designs based on quantum expander codes using machine learning which can be trained on depolarizing channel with arbitratry parameters. 

\subsection{Contributions}

The key contribution in the paper is an efficient quantum encoder and decoder design that aims to improve the error correction capabilities of the quantum codes, and can be used for general noisy channels. The proposed quantum decoder is a neural network which approximates the decoder function, where the input is the syndrome and the output is the noise correction. Since the decoder table is implemented using the neural network, the memory requirements are significantly alleviated. Further, the proposed quantum decoder does not choose only the low weight errors to find the noise pattern to correct, but finds the most likely error to correct which is important with asymmetric errors. Even though a small-set-flip algorithm has been proposed in \cite{leverrier2015quantum} to alleviate the memory challenges of syndrome table, this small-set-flip algorithm is exponential in the weight of generators which limits applicability to large codes. Our proposed decoder further alleviates this and does not require generators to be low-weight. 

The proposed quantum encoder is based on the expander code framework in \cite{leverrier2015quantum}. However, the exact code design does not account for the distance of the code and optimizes the code using a reinforcement learning framework \cite{sutton2018reinforcement}. More precisely, we use a method called Deep Q-network \cite{mnih2015human,osband2016deep}. In this method, the reward, based on logical error, is optimized over the choice of the code parameters. We note that the main challenge in the use of reinforcement learning on the quantum stabilizer codes is that the used classical code must satisfy the dual-containing (or self-orthogonality) constraint\cite{calderbank1998quantum, calderbank1997quantum}, which is hard to verify using the reinforcement learning framework. Thus, we use a construction where such orthogonality conditions are not needed and just the full-rank assumption of the classical parity-check matrix is required.  

The proposed quantum encoder and decoder have been evaluated on both symmetric and asymmetric errors. It can be seen that the proposed designs outperform the existing encoder and decoder designs, and can be used to generate codes with large parameters. 

The rest of the paper is organized as follows. Section \ref{sec:setup} explains the system model with the encoder and decoder blocks. Section \ref{sec:algorithms} explains the proposed quantum encoder and decoder designs. Section \ref{sec:results} presents the evaluations of the proposed approach. Section \ref{sec:concl} concludes the paper with discussion on future work.

\section{System Model and Problem Formulation}\label{sec:setup}
\begin{figure*}[btp]
    \centering
    \begin{tikzpicture}[auto, node distance=2.5cm, >=latex']
	% Drawing the blocks
    \draw
    node [input] (input1) {} 
    node [block, right of=input1, node distance = 2cm] (encode) {Encoder}
    node [block, right of=encode] (channel) {Channel}
        node at (7.25, -1.25) [block, text width=1.95cm, align=center] (syndrome) {Syndrome Measurement}
        node [block, right of=syndrome, text width=1.4cm, node distance = 2.75cm, align=center] (correction) {Error Detection}
    node at (12,0)[block] (multiply) {Recovery}
    node [text width=2cm, right of=multiply, node distance=3.25cm, align = center] (fault) {Fault Tolerant Computing}
	;
    	
    % Joining blocks
    % Commands \draw with options like [->] must be written individually
	\draw[->](input1) -- node {$\ket{b_k}$}(encode);
	\draw[->] (4.5,1.5) -- node {Error, $E_n$}(channel);
	\draw[->](encode) -- node {$\ket{b_n}$}(channel);
	\draw[->](channel) -- node {$\ket{r_n}$}(multiply);
	\draw[->] (5.625,0) |- node {}(syndrome);
	\draw[->](syndrome) -- node {$s_m$}(correction);
	\draw[->, pos=0.3](correction) -| node {$\hat{E}_n$}(multiply);
 	\draw[->](multiply) -- node {$\ket{\hat{b}_n}$} (fault);
    \end{tikzpicture}
    \caption{Block diagram representing a quantum encoder and decoder system.}
    \label{fig:bd_quantum_system}
\end{figure*}
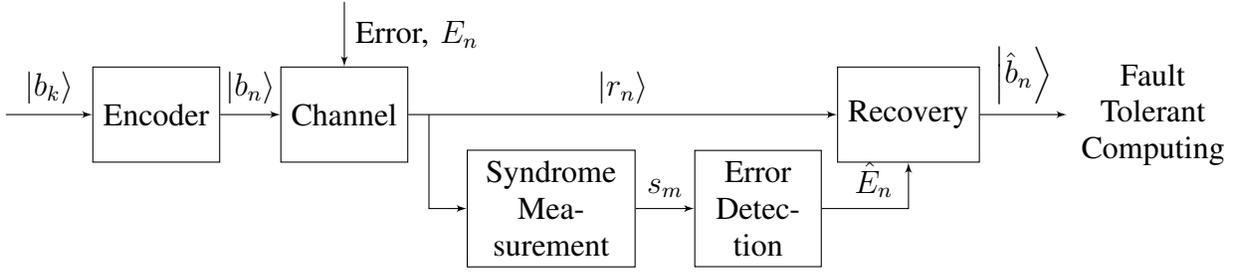

The block diagram for the quantum communication is depicted in Fig. \ref{fig:bd_quantum_system}. Different blocks of the system are explained in the following paragraphs.

\subsection{Encoder}
%\subsubsection{Stabilizer codes}
The Pauli group $G_1$ on single qubit is generated from the Pauli matrices  $\expval{X,Y,Z}$. Pauli group $G_n$ on $n$ qubits is generated by applying the Pauli operators on $n$ qubits and taking their tensor product. The quantum error correcting code used in this paper is an $[[n,k]]$ stabilizer code that encodes $k$ logical qubits into $n$ physical qubits. An $[[n,k]]$ stabilizer code is defined to be the vector space $V_S$ stabilized by a subgroup $S$ of the $n$-fold Pauli group $G_n$. The subgroup $S$ does not contain $-I$, {\em i.e.}, $-I \notin S$ and has $n-k$ independent and commuting generators, $S = \expval{g_1, \ldots, g_{n-k}}$.

The generators of any stabilizer code can be represented using binary vector spaces \cite{gottesman1997stabilizer}. This representation can be used to derive quantum error correcting codes from classical error correcting codes. In this representation, the $n-k$ generators are represented using an $(n-k)\times 2n$ binary matrix with a vertical line dividing the matrix in two halves. The rows of this binary matrix correspond to different generators and columns to different qubits. The first half has a 1 whenever the corresponding generator has $X$ or $Y$ in that qubit position. And the second half has a 1 whenever the corresponding generator has $Z$ or $Y$ in that qubit position. Such a matrix is called the check matrix of the stabilizer code.

%\subsubsection{CSS codes}
Quantum CSS codes are special instances of stabilizer codes that are constructed from classical binary linear codes \cite{calderbank1998quantum,calderbank1997quantum}. For any two classical linear codes $C_X$ and $C_Z$ with parity-check matrices $H_X$ and $H_Z$ with $C_X^\bot \subset C_Z$ and $C_Z^\bot \subset C_X$, the check matrix of the quantum CSS code can be defined as:
\begin{equation}\label{eq:css}
{\cal H} =     \left[
    \begin{array}{c|c}
        H_X & 0\\
        0 & H_Z\\
    \end{array}
    \right].
\end{equation}
The orthogonality of the two classical codes ensure that the commutative property of generators is satisfied. One example of CSS codes is the 7-qubit Steane code, which is generated from the classical [7, 4, 3] Hamming code. %Defining $H_X = H_Z = H$, where $H$ is the parity-check matrix of Hamming code satisfies the properties mentioned above required for constructing CSS code. So the Steane code is generated using (\ref{eq:css}) and setting $H_X = H_Z = H$.

In this paper, we use the code design based on the quantum expander codes in \cite{leverrier2015quantum}, where any given classical linear $[n,k]$ code $C$ with parity-check matrix $H$ (of dimensions $n-k \times n$) can be used to generate a CSS code as: %But for any general linear classical code $C$ with parity-check matrix $H$ for an [n,k] classical linear code, the following expressions can be used to generate two orthogonal parity-check matrices \cite{small_set_flip}.
\begin{equation}\label{eq:hx}
    H_X = \left[I_{n-k}\otimes H, H^T\otimes I_{n}\right],
\end{equation}
\begin{equation}\label{eq:hz}
    H_Z = \left[H\otimes I_{n-k}, I_{n}\otimes H^T\right]
\end{equation}
where, $I_n$ is the $n \times n$ identity matrix. The dimensions of ${\cal H}$ thus becomes $[[n^2+(n-k)^2, k^2]]$. This representation ensures that the classical linear codes generated from $H_X$ and $H_Z$ are orthogonal to each other and hence a valid CSS code can be generated using (\ref{eq:css}). Using the quantum parity check matrix, generators of the quantum code can be found \cite{gottesman1997stabilizer}. 

\subsection{Depolarizing channel}
The channel used in this paper is an asymmetric depolarizing channel.  Most of the quantum computing devices \cite{astafiev2004quantum} are characterized by relaxation times that are one-two orders of magnitude larger than the corresponding dephasing (loss of phase coherence, phase-shifting) times. Relaxation leads to both bit-flip and phase-flip errors, whereas dephasing only leads to phase-flip errors. Such asymmetry  translates to an asymmetry in the occurrence probability of bit-flip  and phase-flip errors. Thus, a general asymmetric depolarizing channel is modeled for each quantum bit as:
\begin{equation*}
\begin{split}
    \rho \xrightarrow{} \varepsilon(\rho) = (1&-p_x-p_y-p_z) \rho + p_X X\rho X \\
                                              &+ p_Y Y\rho Y + p_Z Z\rho Z.
\end{split}
\end{equation*}
This can be interpreted as the state $\rho$ is left alone with probability $1-p_X-p_Y-p_Z$ and the operators $X$, $Y$ and $Z$ are applied with probabilities $p_X$, $p_Y$, and $p_Z$ respectively. We assume that the error on each of the $n$ physical qubits is independent of each other.

%The depolarizing channel is a type of quantum noise that polarizes a qubit with a given  probability $p$. The state of a quantum system $\rho$ passing through this channel will be 

\subsection{Syndrome measurement, error detection and recovery}
To detect the error in received qubits, the syndromes are measured using each generator. The eigenvalues of each generators of the stabilizer are measured to obtain $n-k$ syndromes $\beta_1, \ldots, \beta_{n-k}$. That is, for any error $E_n$, the syndrome $\beta_i$ corresponding to the generator $g_i$ is given by $E_n g_i E_n^H = \beta_i g_i$.

After measuring the syndromes, an error $\hat{E}_n$ is picked corresponding to the measured syndrome and $\hat{E}_n^H$ is applied to the received qubits to achieve recovery.

In this paper, received qubits are in error when after error correction $E_n$ is not equal to $\hat{E}_n$. The error probability thus becomes $\Pr(E_n \neq \hat{E}_n)$. This provides an upper bound on the error probability, because the errors may further be corrected after decoding, since decoding is a conversion to smaller space. Note that the error is counted in the encoded space, this is because further computations in practical systems may be applied on encoded  qubits making the encoded error important \cite{gottesman1997stabilizer}. 
% The reason of counting error in the encoded space is that further computation may be applied in the practical system on encoded qubits making the encoded error more important

\subsection{Problem Formulation}

Having discussed the different components in the quantum communication system, we aim to use machine learning approaches to find the classical encoder $H$ for generation of the quamtum codes as well as the error $\hat{E}_n$ which is picked  corresponding to the measured syndrome. 

%After error correction, the $n$ encoded qubits need not actually be completely decoded back to $k$ qubits. Fault tolerant quantum computing can be done on the data while it is still encoded in the later stages of quantum error correction \cite{gottesman}.

%\section{Performance Improvement as an Optimization Problem}\label{sec:formulation}
%[---Improve this section title---]

%We target to improve the performance of overall system by formulating it as a two-step joint optimization problem. The two parts are the encoder part and the error correcting part as explained in the following sub sections.

\section{Proposed Algorithm}\label{sec:algorithms}

In this section, we present the proposed algorithm. We will first present the proposed quantum decoder using neural network. We will then present the quantum encoder design that is based on reinforcement learning which will use the proposed quantum decoder thus giving a joint encoder and decoder design for the quantum codes. 

\subsection{Quantum Decoder using Deep Neural Network}\label{sec:nn_explain}

The key job of the quantum decoder is to map the syndrome $\bar{s}$ to the error $\hat{E}_n(\bar{s})$ that will be corrected after the channel. In order to learn this mapping, we use a deep neural network as a function approximator and train it with $\bar{s}$ as the input and $\hat{E}_n(\bar{s})$ as the output. In order to generate training examples to train such a network, we need to find the mapping $\hat{E}_n(\bar{s})$ for some syndromes $\bar{s}$. 

Since different errors can lead to same syndrome (being a many-to-one mapping), finding the error is not a straightforward task. In the traditional approach, among the different possible errors corresponding to a syndrome, the error with lowest weight is chosen. However, this may not be the best approach when the errors are asymmetric, since certain errors of low weight may be less likely than other errors of high weight. In order to alleviate that, we use the most likely error rather than the lowest weight error. More precisely, for a quantum parity check matrix ${\cal{H}}$ corresponding to $[[n^2 + k^2,(n-k)^2]]$ quantum code, the  channel error $\hat{E}_n$ for a given syndrome $\bar{s} = \beta_1, \ldots, \beta_{n-k}$ is predicted as
\begin{equation}\label{eq:decoder}
\hat{E}_n(\bar{s}, {\cal{H}}) = \argmax_{E_n} \Pr(E_n \mid \bar{s}(E_n, {\cal{H}}) = \bar{s})
\end{equation}
where, $\bar{s}(E_n, {\cal{H}})$ is the syndrome of channel error $E_n$ for the stabilizer code generated from ${\cal{H}}$. Note that the estimated error $\hat{E}_n(\bar{s}, {{\cal{H}}})$ is function of both the measured syndrome $\bar{s}$ and the quantum code ${\cal{H}}$. We note that the output of the neural network is thresholded to map to $\{I, X, Y, Z\}^n$ to map the error $E_n$ since the neural network is not guaranteed to give discrete errors.

\subsection{Encoder Designs using Deep Reinforcement Learning}\label{sec:pg_explain}
To improve the performance of overall quantum communication system, the quantum generators also need to be optimized. The problem of optimizing the generators can be formulated as an error minimization problem. The check matrix ${\cal{H}}$ of the stabilizer code can be optimized as
\begin{equation}\label{eq:encoder}
    % \hspace{-0.6in}
    {\cal{H}}^* = \argmin_{{\cal{H}}} \mathbb{E}_{E_n}\left[\mathds{1}\left( \hat{E}_n\left(\bar{s}(E_n,{\cal{H}}), {\cal{H}}\right)  \neq E_n\right) \right],
\end{equation}
\if 0
\begin{equation}\label{eq:encoder}
    \quad = \argmin_H \mathbb{E}_{E_n}\left[\mathds{1}( f_H(\bar{s}_H(E_n))  \neq E_n) \right]
\end{equation}
\fi 
where $\mathds{1}(\cdot)$ is the indicator function and the expectation is taken over channel errors $E_n$. This minimizes the 0-1 loss in prediction of channel error. %We employ Deep Q-learning to perform this task using the performance of the neural network based error correction as reward function.

%\subsection{Learning new codes using deep Q-learning}
%Reinforcement learning approaches have been used to improve classical codes \cite{rl_classical} and improvement in performance have been shown for classical linear codes and polar codes.

To find an optimal code as given by the minimizer of (\ref{eq:encoder}), we use deep reinforcement learning algorithm with $-\mathbb{E}_{E_n}\left[\mathds{1}( \hat{E}_n  \neq E_n) \right]$ as reward. That is, we construct new codes by updating the check matrix according to a reward that is evaluated as negative of 0-1 loss of neural network trained to correct errors. Since our code construction problem requires a large state space, the DQN agent is used to learn the codes \cite{mnih2015human,osband2016deep}. 

\begin{figure}[htbp]
	\centering
	\begin{tikzpicture}[auto, node distance=2cm, >=latex']
    	% Drawing the blocks
        \draw
        node [block] (agent) {DQN Agent}
        node [block, below of=agent, text width=2cm, align = center] (decoder) {NN Error Correction}
        node [block, below of=decoder, text width=2cm, align = center] (environment) {Update Parity Check Matrix}
    	;
        	
        % Joining blocks
        % Commands \draw with options like [->] must be written individually
    	\draw[->] (agent.east) -- ++(1.25,0) --node{$a_t$} ++(0,-4) -- (environment.east);	
    	\draw[->] (environment.west) --node{$H_{t+1}$} ++(-1.15,0) -- ++(0,4) --node{$H_{t}$}  (agent.west);	
    	\draw[->] (environment.north) -- node{$H_{t+1}$}  (decoder.south);		
    	\draw[->] (decoder.north) -- node{$r_t$}  (agent.south);	
    \end{tikzpicture}
	\caption{Scheme for learning new codes using deep Q network agent.}
	\label{fig:dqn_scheme}
\end{figure}
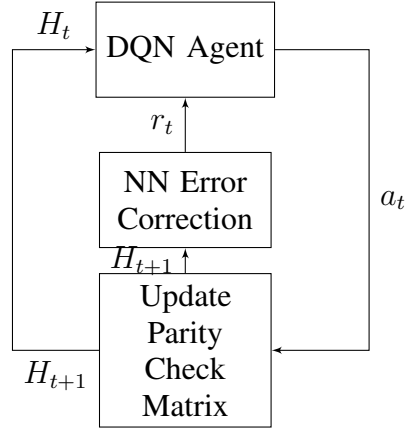

The overall scheme followed for the code construction is shown in Fig. \ref{fig:dqn_scheme}. The different blocks and notations of the scheme in Fig. \ref{fig:dqn_scheme} are explained below.

\begin{enumerate}
	\item State ($H_t$): The parity-check matrix $H_t$ of the classical code is the current state at $t$-th step. This matrix is used to get the quantum stabilizer code as shown in Equations (\ref{eq:hx}), (\ref{eq:hz}) and (\ref{eq:css}). Let the dimensions of this matrix be $n_1 \times n_2$. 
	\item Agent (DQN Agent): The DQN agent takes the current state $H_t$ as the input and suggests an action $a_t$. DQN is explained in detail in the next paragraph.
	\item Action ($a_t$): The DQN agent takes $H_t$ as the input and outputs a discrete action $a_t$. The action $a_t$ is an $n_1 \times n_2$ matrix with all zero entries and one non-zero entry. The position of the single non-zero entry represents the position on the parity-check matrix $H_t$ that needs to be updated. This construction action allows us to avoid the large action space of all possible parity matrices($<2^{n_1\times n_2}$).
	\item Environment (Update Parity-Check Matrix): The parity-check matrix $H_{t}$ is updated by flipping the bit at the position specified by the non-zero entry of the action $a_t$ to get the new parity-check matrix $H_{t+1}$. 
	\item Reward (NN Error Correction): For every action suggested by the DQN agent, CSS code is generated from the parity-check matrix $H_{t+1}$ using (\ref{eq:hx}), (\ref{eq:hz}), and (\ref{eq:css}). This CSS code is simulated for multiple random trials using the neural network based error correction. The performance of this CSS code is evaluated using the 0-1 loss of this neural network. This performance measure is used in part to calculate the reward $r_t$ of DQN agent. 
	\item Reward ($r_t$): The reward is composed of two elements. The first is the  0-1 loss of neural network used for error correction, $-\frac{1}{N}\sum_{E_n}\mathds{1}( \hat{E}_n  \neq E_n)$, where the summation is over multiple random trials with channel error $E_n$, and $\hat{E}_n$ is the channel error estimated by the neural network. The second is the rank-deficiency of the parity-check matrix $H_{t+1}$, since we require this matrix to be full-rank for an efficient choice of the quantum code. Thus, the chosen reward function is 
	
\if 0	
	The 0-1 loss of neural network used for error correction is obtained as 
	\begin{equation*}
	r_t' = -\frac{1}{N}\sum_{E_n}\left[\mathds{1}( \hat{E}_n  \neq E_n) \right]
	\end{equation*}

	For the parity-check matrix $H_{t+1}$ used to generate the quantum code to be a valid classical linear code, it needs to be full row rank. To ensure this, a penalty is added to the reward as 
	\fi
	\begin{equation}\label{eq:reward}
	    	r_t = -\frac{1}{N}\sum_{E_n}\left[\mathds{1}( \hat{E}_n  \neq E_n)\right] + \left(\text{Rank}(H_{t+1}) - n_1\right),
	\end{equation}
	where $n_1$ is the number of rows of $H_{t+1}$. This reward is then returned to the DQN agent.
\end{enumerate}
The DQN agent is run for sufficient number of steps $T$ greater than $n_1\times n_2$ so as to allow for all the bits to flip and the final state $H_T$ is used to construct the new CSS code.

\noindent {\bf Deep Q-Network \cite{mnih2015human,osband2016deep}:}
The goal of DQN agent is to select the actions such that the reward is maximized. DQN considers a future discounted reward at each step $t$ and defines the optimal action-value function $Q^*(H_t,a_t)$ that follows the Bellman equation as
\begin{equation}
    Q^*(H_t,a_t) =  \mathbb{E}_{H_{t+1}} \left[r_t + \gamma \max_{a_{t+1}} Q(H_{t+1}, a_{t+1}) \right]
\end{equation}
where, $\gamma$ is the the future discount factor. We keep $\gamma$ large (close to 1) as we want the final reward to be maximized. The optimal strategy followed by DQN is to select an action $a_t$ that maximizes the action-value function $Q^*(H_t,a_t)$.

A neural network with weights $\theta$ is used to estimate the value of $Q^*(H,a) \approx Q(H,a;\theta)$. Such a neural network is called a Q-network. A Q-network is trained by minimizing a loss function $L_i(\theta_i)$ at each iteration $i$ between predicted Q-values and target Q-values as

\begin{equation}\label{eq:loss_func}
\begin{split}
    L_i(\theta_i) = \frac{1}{|\mathcal{B}|}\sum_{r_k, H_k, a_k, H_{k+1}\in \mathcal{B}}\bigg[ \Big( r_k + \gamma \max_{a'} Q(H_{k+1}, a'; \theta_{i-1}) - Q(H_{k}, a_{k}; \theta_{i}) \Big)^2 \bigg]
\end{split}
\end{equation}
where $\mathcal{B}$ is replay memory consisting of state, action, and reward from past. The loss is optimized using gradient descent. More details on DQN can be found in \cite{mnih2015human,osband2016deep}. Algorithm \ref{algo:main} explains the algorithm used to learn new codes using DQN agent as explained in the scheme of Fig. \ref{fig:dqn_scheme}. 

\begin{algorithm}
	\SetAlgoLined
	
	 Initialize replay memory $\mathcal{D}$ and action-value function
	 
    \For{episode=1:total number of episodes} {
    
    Initialize the state sequence $\{H_1\}$
    
    \For{t=1:T} {
        Choose action $a_t$ such that: $ a_t = \argmax_{a} Q(H_t, a; \theta)$
        
        Update $H_{t}$ by flipping $a_t$-th bit
        
        Simulate the code formed from $H_t$ and calculate the reward $r_t$ using (\ref{eq:reward})
        
        Store the transition $(H_t,a_t,R_t,H_{t+1})$ in replay memory
        
        Sample a mini-batch $\mathcal{B}$ from replay memory $\mathcal{D}$ uniformly randomly
        
        Perform gradient descent step on $L(\theta)$ given in  (\ref{eq:loss_func})
        }
    }
	\caption{Learning New Codes}\label{algo:main}
\end{algorithm}

In Algorithm \ref{algo:main}, the parity-check matrix is initialized and DQN agent is used to get the position of the bit that should be flipped. Performance is measured for the updated check matrix and reward is returned to the DQN agent.

\section{Simulations  Results}\label{sec:results}

In this section, we will describe the simulation results based on the proposed quantum encoder and decoder designs. In order to compare the decoder designs, we use the following two comparable algorithms:

\noindent {\bf 1. Error correction using syndrome lookup table: }
A very simple method to predict channel error from syndromes is to use a syndrome lookup table. A syndrome lookup table can be generated giving higher priorities to lower weight errors and can be looked up to predict the errors. While this method is very simple and fast, it has few disadvantages. One major disadvantage of this method is memory usage. Syndrome lookup tables take up large spaces of memory that increases exponentially with the dimensions of the code making this method unusable for large codes. Another disadvantage of this method is that the lower weight errors are given higher priorities and the three types of errors $X$, $Y$ and $Z$ are given equal priorities. So, this method does not perform well when the depolarizing channel has high probability of error or has asymmetric errors.

\noindent {\bf 2. Small-set-flip algorithm: } An efficient algorithm to predict the channel error from syndromes for quantum expander codes is given in \cite{leverrier2015quantum}. This algorithm goes through all the generators and for each generator, it checks whether flipping any pattern of bits strictly decreases the weight of the sequence of syndromes. Even though this algorithm runs in time linear in the number of qubits, the run time increases exponentially with the weight of the generators. This is because the algorithm involves searching through the errors whose support is included in some generator such that the syndrome weight is maximally reduced. Hence, this algorithm cannot be used efficiently with quantum stabilizer codes where the weight of generators can be large.

The used  deep neural network for evaluations has parameters as given in Table I. We will first present the evaluation results for the quantum decoder, which will then be followed by the results for the joint quantum encoder and decoder design.

% Training data was generated with inputs as syndromes and outputs as channel errors with maximum probability for the given syndromes.

%Unlike the syndrome lookup method, this method of error correction can be used for large codes since there is no problem of memory usage. And this code can also be used for quantum stabilizer codes whose generators have larger weights unlike the small-set-flip algorithm.

\if 0
Simulations were performed in the following manner:
\begin{enumerate}
    \item The neural network based error correction explained in section \ref{sec:nn_explain} was compared against standard error correction algorithms. This simulation was done for the five qubit code and for different channel parameters.
    \item CSS code learnt using DQN as explained in section \ref{sec:pg_explain} was compared against the CSS code generated from (\ref{eq:css}). Since the dimensions of these codes was large, this simulation was done only using the neural network based error correction method. Simulations for different channel parameters were also done.
\end{enumerate} 
These results are discussed in the following sub sections.
\fi 

\begin{center}
	\captionof{table}{Parameters for Deep Neural Network}
	\begin{tabular}[htbp]{c c}
		Parameters & Values \\
		\hline
		\hline
		Number of hidden layers & 5 \\
		Number of neurons per layer & 100 \\
		Number of samples for training & 5000 \\
		Batch size & 100 \\
		Optimizer & Adam \\
		Learning rate & 0.01 \\
		Total number of epochs  & 1000
	\end{tabular}
\end{center}

\subsection{Deep Neural Network Decoder}
\begin{figure*}[htbp]
    \centering
    \subfigure[Channel parameters $p_y = p_z = p_x$.]{
        % This file was created by matplotlib2tikz v0.7.4.
\begin{tikzpicture}
\tikzstyle{every node}=[font=\small]

\begin{axis}[
legend cell align={left},
legend style={at={(0.03,0.97)}, anchor=north west, draw=white!80.0!black},
tick align=outside,
tick pos=left,
x grid style={white!69.01960784313725!black},
xlabel={\(\displaystyle p_x\)},
xmajorgrids,
xmin=-0.015, xmax=0.315,
xtick style={color=black},
xtick={-0.05,0,0.05,0.1,0.15,0.2,0.25,0.3,0.35},
xticklabels={−0.05,0.00,0.05,0.10,0.15,0.20,0.25,0.30,0.35},
y grid style={white!69.01960784313725!black},
ylabel={Error rate},
ymajorgrids,
ymin=-0.04483, ymax=0.94143,
ytick style={color=black},
ytick={-0.2,0,0.2,0.4,0.6,0.8,1},
yticklabels={−0.2,0.0,0.2,0.4,0.6,0.8,1.0}
]
\addplot [semithick, red, mark=*, mark size=3, mark options={solid}]
table {%
0 0
0.0157894736842105 0.01408
0.0315789473684211 0.04728
0.0473684210526316 0.08884
0.0631578947368421 0.14344
0.0789473684210526 0.20256
0.0947368421052632 0.26596
0.110526315789474 0.33864
0.126315789473684 0.3962
0.142105263157895 0.45604
0.157894736842105 0.50484
0.173684210526316 0.55808
0.189473684210526 0.60088
0.205263157894737 0.6464
0.221052631578947 0.68048
0.236842105263158 0.71916
0.252631578947368 0.75556
0.268421052631579 0.79448
0.284210526315789 0.83368
0.3 0.8672
};
\addlegendentry{Syndrome lookup}
\addplot [semithick, green!50.0!black, mark=triangle*, mark size=3, mark options={solid}]
table {%
0 0
0.0157894736842105 0.0602
0.0315789473684211 0.1158
0.0473684210526316 0.1658
0.0631578947368421 0.2092
0.0789473684210526 0.267
0.0947368421052632 0.3148
0.110526315789474 0.3658
0.126315789473684 0.4066
0.142105263157895 0.4568
0.157894736842105 0.502
0.173684210526316 0.5364
0.189473684210526 0.5904
0.205263157894737 0.631
0.221052631578947 0.6784
0.236842105263158 0.7142
0.252631578947368 0.7526
0.268421052631579 0.8008
0.284210526315789 0.8468
0.3 0.8966
};
\addlegendentry{Small-set-flip}
\addplot [semithick, blue, mark=square*, mark size=2.1, mark options={solid}]
table {%
0 0
0.0157894736842105 0.03964
0.0315789473684211 0.0856
0.0473684210526316 0.13744
0.0631578947368421 0.19596
0.0789473684210526 0.26192
0.0947368421052632 0.31604
0.110526315789474 0.37084
0.126315789473684 0.42728
0.142105263157895 0.4742
0.157894736842105 0.5292
0.173684210526316 0.57392
0.189473684210526 0.60632
0.205263157894737 0.65488
0.221052631578947 0.70084
0.236842105263158 0.72964
0.252631578947368 0.74856
0.268421052631579 0.73352
0.284210526315789 0.7184
0.3 0.70336
};
\addlegendentry{Neural network}
\end{axis}

\end{tikzpicture}
        \label{fig:decode_plot_1}
    }\subfigure[Channel parameters $p_y = 0.05 p_x$ and $p_z = 0.05 p_x$.]{
        % This file was created by matplotlib2tikz v0.7.4.
\begin{tikzpicture}
\tikzstyle{every node}=[font=\small]

\begin{axis}[
legend cell align={left},
legend style={at={(0.03,0.97)}, anchor=north west, draw=white!80.0!black},
tick align=outside,
tick pos=left,
x grid style={white!69.01960784313725!black},
xlabel={\(\displaystyle p_x\)},
xmajorgrids,
xmin=-0.015, xmax=0.315,
xtick style={color=black},
xtick={-0.05,0,0.05,0.1,0.15,0.2,0.25,0.3,0.35},
xticklabels={−0.05,0.00,0.05,0.10,0.15,0.20,0.25,0.30,0.35},
y grid style={white!69.01960784313725!black},
ylabel={Error rate},
ymajorgrids,
ymin=-0.0178, ymax=0.3738,
ytick style={color=black},
ytick={-0.05,0,0.05,0.1,0.15,0.2,0.25,0.3,0.35,0.4},
yticklabels={−0.05,0.00,0.05,0.10,0.15,0.20,0.25,0.30,0.35,0.40}
]
\addplot [semithick, red, mark=*, mark size=3, mark options={solid}]
table {%
0 0
0.0157894736842105 0.00168
0.0315789473684211 0.00708
0.0473684210526316 0.01304
0.0631578947368421 0.02296
0.0789473684210526 0.03904
0.0947368421052632 0.05168
0.110526315789474 0.07504
0.126315789473684 0.08664
0.142105263157895 0.10636
0.157894736842105 0.12576
0.173684210526316 0.1486
0.189473684210526 0.1712
0.205263157894737 0.19616
0.221052631578947 0.2212
0.236842105263158 0.23596
0.252631578947368 0.2588
0.268421052631579 0.29056
0.284210526315789 0.31188
0.3 0.33928
};
\addlegendentry{Syndrome lookup}
\addplot [semithick, green!50.0!black, mark=triangle*, mark size=3, mark options={solid}]
table {%
0 0
0.0157894736842105 0.0204
0.0315789473684211 0.0434
0.0473684210526316 0.0528
0.0631578947368421 0.0682
0.0789473684210526 0.0952
0.0947368421052632 0.12
0.110526315789474 0.1316
0.126315789473684 0.1524
0.142105263157895 0.1762
0.157894736842105 0.1814
0.173684210526316 0.1992
0.189473684210526 0.2246
0.205263157894737 0.2484
0.221052631578947 0.2638
0.236842105263158 0.2732
0.252631578947368 0.3052
0.268421052631579 0.3152
0.284210526315789 0.346
0.3 0.356
};
\addlegendentry{Small-set-flip}
\addplot [semithick, blue, mark=square*, mark size=2.1, mark options={solid}]
table {%
0 0
0.0157894736842105 0.00292
0.0315789473684211 0.0084
0.0473684210526316 0.0164
0.0631578947368421 0.02428
0.0789473684210526 0.02892
0.0947368421052632 0.03284
0.110526315789474 0.04176
0.126315789473684 0.05476
0.142105263157895 0.06608
0.157894736842105 0.07412
0.173684210526316 0.087
0.189473684210526 0.09788
0.205263157894737 0.10884
0.221052631578947 0.13528
0.236842105263158 0.14944
0.252631578947368 0.16984
0.268421052631579 0.19756
0.284210526315789 0.21772
0.3 0.234
};
\addlegendentry{Neural network}
\end{axis}

\end{tikzpicture}
        \label{fig:decode_plot_2}
    }
    \caption{Error rate comparison between different error correction methods for [[5,1]] five qubit code.}
    \label{fig:decode_plot}
\end{figure*}
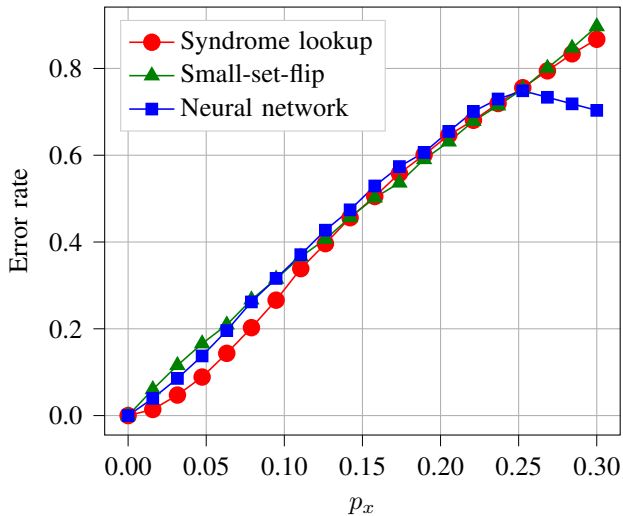
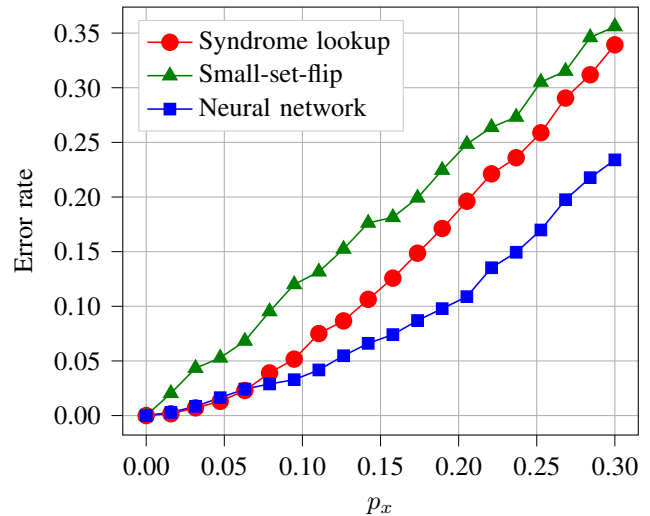

In this subsection, we evaluate the performance of the proposed quantum decoder. The details of the parameters used for the deep neural network are given in Table I.  In order to see the improved performance, we  used the five qubit code \cite{PhysRevA.54.3824}, which is the shortest code
that can protect against depolarizing one-qubit errors. Fig. \ref{fig:decode_plot} shows the  loss error rate, $\Pr(\hat{E}_n\ne E_n)$, that occurred during the error correction using lookup table, small-set-flip algorithm, and the proposed deep neural network based methods. Fig. \ref{fig:decode_plot_1} shows the results for a depolarizing channel with equal probabilities for $X$, $Y$ and $Z$ errors, $p_x = p_y = p_z$. In the symmetric case of Fig. \ref{fig:decode_plot_1} where $p_x = p_y = p_z$, the performance of all the three methods is almost the same. But for lower probabilities, lookup table performs slightly better than the neural network based method since the lookup table gives higher priorities to lower weight errors. The proposed algorithm outperforms small-set-flip algorithm for low error probabilities depicting that we can gain accuracy while at the same time reducing complexity of the algorithm. Further, at higher errors, lower weight errors are not more likely and thus the proposed algorithm outperforms both the baseline algorithms.  

Fig. \ref{fig:decode_plot_2} shows the result for an asymmetric channel where probabilities of $Y$ and $Z$ errors are smaller than that of $X$ error, $p_y = p_z = 0.05p_x$. The improvement in performance of error correction using deep neural network can be seen for asymmetric channels as shown in Fig. \ref{fig:decode_plot_2}. The syndrome lookup table method and the small-set-flip algorithm inherently consider $X$, $Y$ and $Z$ errors to have equal probabilities, meanwhile the neural network learns to correct the errors according to their asymmetric distribution. Hence, a significant improvement can be seen in the error correction performance. This depicts that the proposed decoder outperforms the baseline algorithms for asymmetric error probabilities. 

%Also, as explained in section \ref{sec:nn_error_explain}, this neural network based method can be used for any general stabilizer codes with larger dimensions. Hence this method is employed in the construction of new codes using deep Q-learning.

\subsection{Joint Quantum Encoder and Decoder}

In this subsection, we will evaluate the performance of the proposed encoder and decoder design, where the encoder is based on the deep reinforcement learning and uses the trained deep neural network decoder in each iteration. The parameters for the decoder remains same as in Table I, while the parameters for the deep reinforcement learning are chosen as in Table II.  In order to perform the evaluations, $[[58,16]]$ (from $[n,k]=[7,4]$) code was learnt. For every parity-check matrix suggested by DQN agent, the training examples are generated and the decoder neural network is trained for error correction. The 0-1 loss performance of the neural network is used as the reward to learn new codes. This code can be compared with the  standard $[[58,16]]$ CSS code generated using (\ref{eq:hx}) and (\ref{eq:hz}) where $H$ is the Hamming code. For the learnt code, $H$ is taken as the variable parameters which will be learnt using reinforcement learning. Since the dimensions of these codes was large, we did not use the syndrome lookup as a baseline.

\begin{center}
	\captionof{table}{Parameters for Deep Reinforcement Learning}
	\begin{tabular}{c c}
		Parameters & Values \\
		\hline
		\hline
		Number of steps per episode & 32 \\
		Future discount factor & 0.99 \\
		Policy & Epsilon greedy \\
		Optimizer & Adam \\
		Learning rate & 0.001 \\
	\end{tabular}
\end{center}

%Using the error rate performance of the neural network as reward for the DQN agent, new CSS code with dimensions [[58,16]] was learnt. This code was compared against the standard [[58,16]] CSS code generated using (\ref{eq:hx}) and (\ref{eq:hz}) with $H$ as the Hamming code. 

\begin{figure*}[htbp]
    \centering
    \subfigure[Channel parameters $p_y = p_z = p_x$.]{
        % This file was created by matplotlib2tikz v0.7.4.
\begin{tikzpicture}
\tikzstyle{every node}=[font=\small]
\begin{axis}[
legend cell align={left},
legend style={at={(0.03,0.97)}, anchor=north west, draw=white!80.0!black},
tick align=outside,
tick pos=left,
x grid style={white!69.01960784313725!black},
xlabel={\(\displaystyle p_x\)},
xmajorgrids,
xmin=-0.015, xmax=0.315,
xminorgrids,
xtick style={color=black},
xtick={-0.05,0,0.05,0.1,0.15,0.2,0.25,0.3,0.35},
xticklabels={−0.05,0.00,0.05,0.10,0.15,0.20,0.25,0.30,0.35},
y grid style={white!69.01960784313725!black},
ylabel={Error rate},
ymajorgrids,
ymin=-0.045984655, ymax=0.965677755,
yminorgrids,
ytick style={color=black},
ytick={-0.2,0,0.2,0.4,0.6,0.8,1},
yticklabels={−0.2,0.0,0.2,0.4,0.6,0.8,1.0}
]
\addplot [semithick, red, mark=*, mark size=3, mark options={solid}]
table {%
0 0
0.0157894736842105 0.04712414
0.0315789473684211 0.0951931
0.0473684210526316 0.14165172
0.0631578947368421 0.18951379
0.0789473684210526 0.23557586
0.0947368421052632 0.28446552
0.110526315789474 0.3309069
0.126315789473684 0.37913793
0.142105263157895 0.42772759
0.157894736842105 0.49343448
0.173684210526316 0.54207586
0.189473684210526 0.58911379
0.205263157894737 0.64513103
0.221052631578947 0.69231724
0.236842105263158 0.74104828
0.252631578947368 0.77799655
0.268421052631579 0.82547241
0.284210526315789 0.8725069
0.3 0.9196931
};
\addlegendentry{CSS code, Small-set-flip}
\addplot [semithick, green!50.0!black, mark=triangle*, mark size=3, mark options={solid}]
table {%
0 0
0.0157894736842105 0.04227586206896552 
0.0315789473684211 0.0926551724137931 
0.0473684210526316 0.1406896551724138 
0.0631578947368421 0.19389655172413794 
0.0789473684210526 0.2403103448275862 
0.0947368421052632 0.2856551724137931 
0.110526315789474 0.33727586206896554 
0.126315789473684 0.3817931034482759 
0.142105263157895 0.43058620689655175 
0.157894736842105 0.4794827586206897 
0.173684210526316 0.5292413793103449 
0.189473684210526 0.5898275862068966 
0.205263157894737 0.6483448275862069 
0.221052631578947 0.7035862068965517 
0.236842105263158 0.7405172413793103 
0.252631578947368 0.7496551724137931 
0.268421052631579 0.7374137931034482 
0.284210526315789 0.7168620689655173 
0.3 0.7012068965517242 
};
\addlegendentry{CSS code, Neural network}
\addplot [semithick, blue, mark=square*, mark size=2.5, mark options={solid}]
table {%
0 0
0.0157894736842105 0.03796551724137931 
0.0315789473684211 0.08510344827586207 
0.0473684210526316 0.1269655172413793 
0.0631578947368421 0.17396551724137932 
0.0789473684210526 0.21437931034482757 
0.0947368421052632 0.2629655172413793 
0.110526315789474 0.3062758620689655 
0.126315789473684 0.34920689655172416 
0.142105263157895 0.3972413793103448 
0.157894736842105 0.44375862068965516 
0.173684210526316 0.4839310344827586 
0.189473684210526 0.5365172413793103 
0.205263157894737 0.579103448275862 
0.221052631578947 0.641370465068067 
0.236842105263158 0.6801724137931034 
0.252631578947368 0.6973448275862069 
0.268421052631579 0.6868965517241379 
0.284210526315789 0.6620689655172414 
0.3 0.6527586206896552 
};
\addlegendentry{Learnt code, Neural network}
 \end{axis}

\end{tikzpicture}
        \label{fig:encode_plot_1}
    }
    \subfigure[Channel parameters $p_y = 0.1 p_x$ and $p_z = 0.1 p_x$.]{
        % This file was created by matplotlib2tikz v0.7.4.
\begin{tikzpicture}

\tikzstyle{every node}=[font=\small]
\begin{axis}[
legend cell align={left},
legend style={at={(0.03,0.97)}, anchor=north west, draw=white!80.0!black},
tick align=outside,
tick pos=left,
x grid style={white!69.01960784313725!black},
xlabel={\(\displaystyle p_x\)},
xmajorgrids,
xmin=-0.015, xmax=0.315,
xminorgrids,
xtick style={color=black},
xtick={-0.05,0,0.05,0.1,0.15,0.2,0.25,0.3,0.35},
xticklabels={−0.05,0.00,0.05,0.10,0.15,0.20,0.25,0.30,0.35},
y grid style={white!69.01960784313725!black},
ylabel={Error rate},
ymajorgrids,
ymin=-0.01876, ymax=0.39396,
yminorgrids,
ytick style={color=black},
ytick={-0.05,0,0.05,0.1,0.15,0.2,0.25,0.3,0.35,0.4},
yticklabels={−0.05,0.00,0.05,0.10,0.15,0.20,0.25,0.30,0.35,0.40}
]
\addplot [semithick, red, mark=*, mark size=3, mark options={solid}]
table {%
0 0
0.0157894736842105 0.01890344827586207 
0.0315789473684211 0.038251724137931035 
0.0473684210526316 0.05654137931034483 
0.0631578947368421 0.07546896551724137 
0.0789473684210526 0.09478275862068966 
0.0947368421052632 0.11322413793103449 
0.110526315789474 0.13259310344827585 
0.126315789473684 0.1509344827586207 
0.142105263157895 0.17629655
0.157894736842105 0.1888931
0.173684210526316 0.21453793
0.189473684210526 0.23197241
0.205263157894737 0.25325517
0.221052631578947 0.26896552
0.236842105263158 0.29004828
0.252631578947368 0.3071
0.268421052631579 0.32787586
0.284210526315789 0.34881379
0.3 0.3752
};
\addlegendentry{CSS code, Small-set-flip}
\addplot [semithick, green!50.0!black, mark=triangle*, mark size=3, mark options={solid}]
table {%
0 0
0.0157894736842105 0.01932759
0.0315789473684211 0.027103448275862068 
0.0473684210526316 0.04444827586206897 
0.0631578947368421 0.06962068965517242 
0.0789473684210526 0.09217241379310345 
0.0947368421052632 0.11255172413793103 
0.110526315789474 0.13403448275862068 
0.126315789473684 0.15072413793103448 
0.142105263157895 0.17282758620689656 
0.157894736842105 0.18855172413793103 
0.173684210526316 0.20455172413793105 
0.189473684210526 0.2259655172413793 
0.205263157894737 0.24544827586206897 
0.221052631578947 0.25641379310344825 
0.236842105263158 0.2796896551724138 
0.252631578947368 0.3017586206896552 
0.268421052631579 0.31675862068965516 
0.284210526315789 0.33827586206896554 
0.3 0.3543103448275862 
};
\addlegendentry{CSS code, Neural network}
\addplot [semithick, blue, mark=square*, mark size=2.1, mark options={solid}]
table {%
0 0
0.0157894736842105 0.013758620689655172 
0.0315789473684211 0.027482758620689655 
0.0473684210526316 0.03796551724137931 
0.0631578947368421 0.05344827586206897 
0.0789473684210526 0.07055172413793104 
0.0947368421052632 0.08272413793103449 
0.110526315789474 0.10386206896551724 
0.126315789473684 0.1237007808720892 
0.142105263157895 0.13710344827586207 
0.157894736842105 0.15293103448275863 
0.173684210526316 0.16779310344827586 
0.189473684210526 0.18475862068965518 
0.205263157894737 0.19986206896551725 
0.221052631578947 0.2130689655172414 
0.236842105263158 0.2266896551724138 
0.252631578947368 0.2486551724137931 
0.268421052631579 0.2600344827586207 
0.284210526315789 0.2765862068965517 
0.3 0.29420689655172416 
};
\addlegendentry{Learnt code, Neural network}
\end{axis}

\end{tikzpicture}
        \label{fig:encode_plot_2}
    }
    \caption{Error rate comparison for [[58,16]] CSS code generated from Hamming code and [[58,16]] code learnt using DQN. Both small-set-flip algorithm and Neural network based error correction are shown.}
    \label{fig:encode_plot}
\end{figure*}
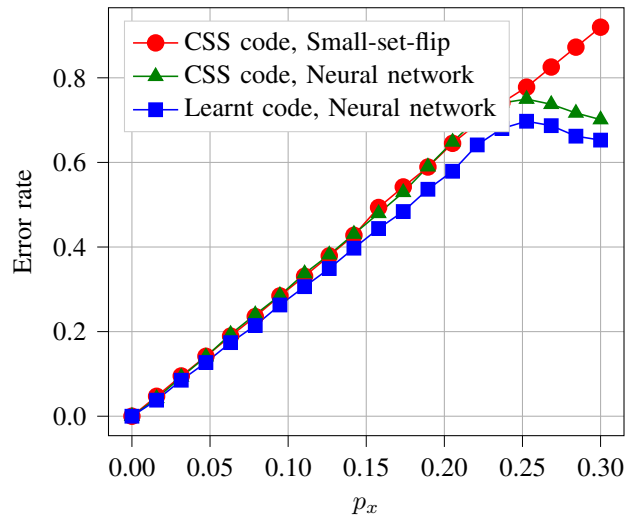
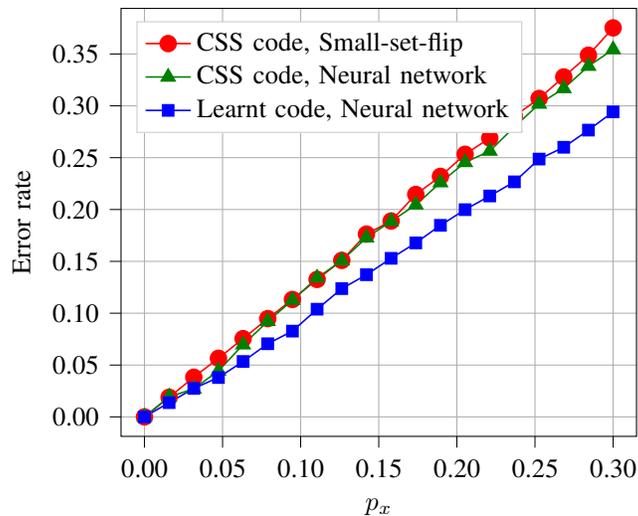

Fig. \ref{fig:encode_plot} shows the error correction performance for the learnt code and the Hamming-based CSS code with error correction done using small-set-flip algorithm and using the trained neural network. Fig. \ref{fig:encode_plot_1} shows the results for a symmetric depolarizing channel with errors $X$, $Y$ and $Z$ having equal probabilities. Fig. \ref{fig:encode_plot_2} shows the results for an asymmetric channel with probabilities $p_y = p_z = 0.1p_x$.

We first note that the proposed deep neural network decoder performs similar (slightly better) than the small-set flip algorithm, while the run-time complexity being the evaluation of neural network (since training was offline and done only once, it is not involved in the run-time). Further, the addition of the encoder design provides significant improvement in performance. The performance improvement is much larger in the asymmetric error probability case. Thus, the decoder alone is not sufficient to outperform the communication systems, and joint encoder and decoder designs are essential.  This demonstrates the applicability of the proposed method for large codes, and for general errors. 

%Error correction performance of the code constructed using DQN is as good as the CSS code with both small-set-flip algorithm and neural network based error correction. Significant improvement in the performance of learnt code can be seen for asymmetric channels as the agent learns to generate codes that result in distinct syndromes for errors with same weight but different error type.
\section{Conclusions and Future Work}\label{sec:concl}

This paper provides novel quantum encoder and decoder designs using machine learning approaches. The quantum decoder is based on deep neural networks used as function approximators, trained using the most likely error for a given syndrome. The quantum encoder uses a CSS based code design, in which a classical code is used to generate the quantum code. This classical code is optimized using a deep reinforcement learning approach, using the proposed quantum decoder. The proposed joint encoder and decoder designs demonstrate significant improvement over the standard code designs for asymmetric channels. Further, the proposed decoder after training has low computational and memory requirements and can be used to replace traditional decoders for quantum codes.

This paper uses a subset of CSS codes, where the quantum code is designed from a classical code. This comes with a constraint on the classical parity check matrix that it needs to be full rank. In order to generalize the algorithm for general quantum codes, there are several other conditions that need to be satisfied \cite{aggarwal2008boolean}. It remains to be investigated if the reinforcement learning based framework can be extended to take the different constraints into account and investigate general stabilizer codes. This will expand the choice of the codes and can further improve the performance of the quantum code. 


% \section*{Acknowledgment}

% The authors would like to thank... GoD



%\bibliographystyle{IEEEtran}
\bibliographystyle{IEEEtran}
\bibliography{ref}
\if 0

\fi


\end{document}